\documentstyle[12pt,a4]{article}
\begin{document}
\begin{titlepage}
\vskip1in
\begin{center}
{\Large {\bf The Super-Liouville Equation on the Half-Line}}
\end{center}
\vskip1in
\begin{center}
{\large
Jo\~ao N.G.N.Prata

Department of Mathematical Sciences

University of Durham

South Road

Durham, DH1 3LE, England}

{\it j.n.g.n.prata@durham.ac.uk}
\end{center}
\vskip1in
\begin{abstract}
A recursive formula for an infinity of integrals of motion for the 
super-Liouville theory is derived. The integrable boundary interactions
for this theory and the super-Toda theory based on the affine
superalgebra $B^{(1)} (0,1)$ are computed. In the first case the boundary
interactions are unambiguously determined by supersymmetry, whilst in
the latter case there are free parameters.
\end{abstract}
\end{titlepage}

\section{Introduction}

The study of boundary quantum integrable models has a wide range of
applications, notably open string theory and dissipative quantum
mechanics \cite{Bowcock} \cite{Schulze}. Considerable progress has
been made with the pioneering work of Ghoshal and Zamolodchikov
\cite{Ghoshal}, where they computed the boundary S-matrix for the
sine-Gordon theory.

In the past few years, there has been renewed interest in the problem of
incorporating fermions in Toda field theories, \cite{Evans} -
\cite{Sorokin}. With the exceptions of the Liouville and sinh-Gordon
theories supersymmetrizing bosonic Toda models is not a simple
matter. If one focusses attention on the integrability of the models,
rather than supersymmetry, it is possible to construct a new class of
Toda models with fermions, where the underlying algebra is a Lie
superalgebra, \cite{Kac} \cite{Olshanetsky}.
Following the work of Zamolodchikov \cite{Zamolodchikov1}
\cite{Zamolodchikov2} some attempts were made at determining exact
S-matrices for this class of theories \cite{Delius} \cite{Grisaru}.

Recently, Inami et al. \cite{Inami} considered the supersymmetric 
extension of the sine-Gordon theory on the half-line and found 
that the requirements of integrability and supersymmetry fully 
determine the boundary potential up to an overall sign.

In this paper, we apply similar considerations to the super-Liouville
theory and the Toda theory based on the superalgebra $B^{(1)} (0,1)$.

Besides its applications in statistical mechanics, the super-Liouville
equation (SLE) arises in Polyakov's approach to the superstring,
\cite{Mansfield}. As a conformal field theory, the SLE can be
described by its integrable structure along with its more conventional
characterization in terms of the Virasoro algebra and its
representations (e.g. \cite{Bazhanov} and references therein). In
fact, an infinite set of involutive integrals of motion (IM) can be
shown to exist. The IM are just composite fields of the stress tensor,
the supercurrents and their derivatives. The theory is then 
characterized by the massless states diagonalizing the IM and their 
factorizable S-matrix. We will show that these IM can be derived using
Lax pair techniques. The boundary conditions will be determined by 
requiring the preservation of the superconformal invariance, 
\cite{Ghoshal} \cite{Cardy} \cite{Schulze}. As in the case of the 
super-sine-Gordon theory \cite{Inami}, this strongly restricts the 
boundary equations of motion, so that there will be no free parameters 
whatsoever. Furthermore, these conditions equally allow for the 
conservation of half of the IM, making it still possible to approach 
the theory on the half-line from the point of view of the boundary 
massless S-matrix.

The $B^{(1)} (0,1)$ theory on the other hand is massive and the
determination of the boundary potential will rely on the premiss
that its form can be conjectured with generality by preserving certain
combinations of the lower spin charges, \cite{Ghoshal}
\cite{Inami}. However, this theory is not supersymmetric and there
will thus be free parameters in the boundary potential.

This paper is organised as follows. In section 2 we define the
theories in the bulk. We discuss superconformal invariance and derive
the IM for the SLE. We also compute the spin 1 and 3 densities of the
$B^{(1)} (0,1)$ theory. In section 3 we establish the boundary
conditions for both theories. Finally, we summarize our results in
section 4.
 
\section{The $B(0,1)$ and $B^{(1)} (0,1)$ theories}

In this section, we define the theories in the bulk. For the
SLE, we discuss supersymmetry and determine a recursive formula for
the IM using a method developed in refs. \cite{Ferrara},
\cite{Girardello}. This formula is a supersymmetrized version of the
IM of ref. \cite{Bazhanov} to which it reduces in the bosonic limit. 
We also derive the spin 1 and 3 conserved densities
for the $B^{(1)} (0,1)$ theory by considering the most general
Ansatz. Bearing in mind that our discussion is strictly classical, the
above ingredients should suffice to construct the integrable boundary
interactions with considerable generality.

Firstly, we establish our notation. Consider two-dimensional
superspace, with units such that $\hbar = c = 1$ and $[mass]=1$, and
the superspace coordinate
$$
Z=(x_{\mu}, \theta_A ) = ( x_0, x_1, \theta_1, \theta_2 ),
$$
where $x_{\mu}$ is the coordinate on two-dimensional Minkowski space
and $\theta_A$ are Grassmann variables. We introduce the scalar
superfield $\Phi$ with components:
$$
\Phi = \varphi + i \theta_1 \psi_2 - i \theta_2 \psi_1 + \theta_1 
\theta_2 F.
$$
The superderivatives
$$
D_1= - \partial_{\theta_2} + i \theta_2 \partial_+, \qquad  
D_2= \partial_{\theta_1} + i \theta_1 \partial_-
$$
have the properties
$$
D_1^2=-i \partial_+, \qquad D_2^2 = i \partial_-, 
$$
where the light-cone variables are defined as
$$
x_{\pm}\equiv \frac{x_1 \pm x_0}{2}.
$$

\subsection{The super-Liouville theory}

Let us define the following linear system:
\begin{equation}
\left\{
\begin{array}{l}
D_1 \chi=A_1( \lambda) \chi\\
D_2 \chi=A_2( \lambda) \chi
\end{array}
\right.
\end{equation}
$\chi$ is a column vector, whose components are the bosonic
superfields $V_1$, $V_2$ and the fermionic superfield $V_3$; $\lambda$
is an arbitrary parameter with dimension of mass, and $A_1$, $A_2$ are
the graded matrices:
$$
A_1(\lambda) = - \sqrt{\frac{2}{\lambda}} 
\left(
\begin{array}{c c c}
0 & 0 & e^{2 \Phi}\\
0 & 0 & i e^{2 \Phi}\\
e^{2 \Phi} & i e^{2 \Phi} & 0
\end{array}
\right);
\qquad
A_2(\lambda) = 
\left(
\begin{array}{c c c}
\lambda \theta_1 & -2iD_2 \Phi & 0\\
2iD_2 \Phi & \lambda \theta_1 & - \sqrt{2 \lambda}\\
0 & \sqrt{2 \lambda} & \lambda \theta_1
\end{array}
\right).
$$
The integrability condition for the system (1) is just the $N=1$ SLE,
\begin{equation}
D_1D_2 \Phi = i e^{2 \Phi},
\end{equation}
which is the simplest example of a Toda theory based on a
contragradient Lie superalgebra. This superalgebra is labelled $B(0,1)$
in the calssification of Kac \cite{Kac} and it possesses three
bosonic generators and two fermionic ones. A realization of
$B (0,1)$ is provided by $Osp (1|2;{\cal C})$. The theory based
on this finite superalgebra is conformally invariant. Furthermore, 
the SLE also happens to be supersymmetric and therefore superconformal,
\cite{Mansfield}.

Notice that eq.(2) is independent of the spectral parameter
$\lambda$. This will give rise to an infinity of conservation laws.

Writing out eq.(2) in components, we have:
\begin{equation}
\left\{
\begin{array}{l}
F=-ie^{2 \varphi}; \qquad \partial_- \psi_1 = -2 e^{2 \varphi} \psi_2; 
\qquad \partial_+ \psi_2 = -2 e^{2 \varphi} \psi_1\\
\\
\partial_+ \partial_- \varphi = 2 e^{2 \varphi} ( e^{2 \varphi} + 
2i \psi_1 \psi_2)
\end{array}
\right.
\end{equation}
The above equations of motion can be derived from the superspace action
$$
S=\frac{1}{2} \int d^2 z d^2 \theta (D_1 \Phi D_2 \Phi + i e^{2
\Phi}).
$$
We now define two new scalar superfields U, Z and a fermionic
superfield Y as:
$$
U=lnV_1 + i \lambda x_-; \qquad Z= \frac{V_2}{V_1}; \qquad 
Y= \frac{V_3}{V_1}.
$$
We then have:
\begin{equation}
\begin{array}{l}
D_1U = - \sqrt{\frac{2}{\lambda}} e^{2 \Phi} Y; \qquad 
D_2 U = -2iD_2 \Phi \cdot Z\\
\\
\sqrt{2 \lambda} Y = 2i D_2 \Phi - D_2 Z + 2i D_2 \Phi \cdot Z \cdot Z\\
\\
\sqrt{2 \lambda} Z = D_2 Y - 2i D_2 \Phi \cdot Z \cdot Y
\end{array}
\end{equation}
Taking into account that $Y^2= (D_2 \Phi)^2=0$, we get the following
differential equation for Y:
\begin{equation}
\sqrt{2 \lambda} Y = 2i D_2 \Phi - \frac{i}{\sqrt{2 \lambda}}
\partial_- Y - \frac{1}{\lambda} \partial_- \Phi \cdot D_2 Y \cdot Y.
\end{equation}
We assume an expansion of Y in a power series of $\lambda^{-1}$:
\begin{equation}
Y=\frac{1}{i \sqrt{2 \lambda}} \sum_{n=0}^{\infty} \frac{Y^{(n+1/2)}}{
(2i \lambda)^n}.
\end{equation}
Substituting this expansion in eq.(5) and equating powers of 
$\lambda^{-1}$, we obtain the following recursive formula:
\begin{equation}
\begin{array}{l}
Y^{(1/2)} = -2 D_2 \Phi;\\
Y^{(n+1/2)} = \partial_- Y^{(n-1/2)} - 2i \partial_- \Phi \cdot 
\sum_{l=1}^{n-1} D_2 Y^{(l-1/2)} \cdot Y^{(n-l-1/2)}; \\
n=1,2,3, \cdots
\end{array}
\end{equation}
The integrability condition,
$$ 
D_1 D_2 U = - D_2 D_1 U,
$$
can be interpreted as an infinite number of supersymmetric covariant
conservation laws:
\begin{equation}
D_1 J_2^{(n+1/2)} = D_2 J_1^{(n+1/2)}; \qquad n=1,2,3, \cdots ,
\end{equation}
where
\begin{equation}
\left\{
\begin{array}{l}
J_1^{(n+1/2)} = - e^{2 \Phi} \cdot Y^{(n-1/2)}\\
J_2^{(n+1/2)} = i D_2 \Phi \cdot D_2 Y^{(n-1/2)}
\end{array}
\right.
\end{equation}
One can check that the bosonic conserved quantities will be given by
the $ \theta_1 \theta_2$ component of eq.(8).
We will henceforth work in Euclidean space,
$$
\left\{
\begin{array}{l}
x = x_1\\
 y = i x_0
\end{array}
\right.
\qquad \left\{
\begin{array}{l}
z=x+iy\\
\bar z = x-iy
\end{array}
\right.
$$
and redefine the fields,
$$
\varphi = \phi/2, \qquad \psi_1 = \alpha \bar{\psi}, \qquad \psi_2 =
\alpha \psi,
$$
for future convenience. The parameter $\alpha$ is such that $\alpha^2
= i/2$. The eqsuations of motion (3) then become 
\begin{equation}
\left\{
\begin{array}{l}
F=-ie^{ \phi}; \qquad \partial_z \bar{\psi} = - e^{\phi} \psi; 
\qquad \partial_{\bar z} \psi = - e^{\phi} \bar{\psi}\\
\partial_z \partial_{\bar z} \phi = e^{2 \phi} - e^{\phi} \bar{\psi} \psi
\end{array}
\right.
\end{equation}
The bosonic conservation laws are expressed in the form:
$$
\partial_{\bar z} T_{s+1} = \partial_z \Theta_{s-1}; \qquad
s=1,3,5,\cdots,
$$
where s is the spin of the conserved charge. Here are some elements of
this sequence, which will be useful later:
$$
\begin{array}{l l}
T_2 = & (\partial_z \phi)^2 - \partial_z \psi \psi\\
& \\
T_4 = & (\partial_z^2 \phi)^2 + (\partial_z \phi)^4 + 3(\partial_z
\phi)^2 \psi \partial_z \psi + \partial_z \psi \partial_z^2 \psi\\
& \\
T_6 = & (\partial_z \phi)^6 - \frac{1}{2} \partial_z^2 \phi
\partial_z^4 \phi - \frac{11}{2} (\partial_z \phi)^2
(\partial_z^2 \phi)^2 - \frac{7}{2} (\partial_z \phi)^3 \partial_z^3 
\phi + ( \partial_z \phi)^2 \partial_z^2 \psi \partial_z \psi + \\ 
& \\
& + 8 \partial_z \phi \partial_z^3 \phi \partial_z \psi \psi - 10
\partial_z \phi \partial_z^2 \phi \psi \partial_z^2 \psi - \frac{7}{2}
(\partial_z \phi)^2 \psi \partial_z^3 \psi + 5 (\partial_z \phi)^4 
\psi \partial_z \psi + \\
& \\
& + \frac{1}{2} \partial_z^4 \psi \partial_z \psi - \frac{11}{2}
(\partial_z^2 \phi)^2 \psi \partial_z \psi\\
& \\
& \\
\Theta_0 = & e^{2 \phi} - e^{\phi} \bar{\psi} \psi\\
& \\
\Theta_2 = & 2 (\partial_z \phi)^2 e^{2 \phi} + e^{2 \phi} \psi
\partial_z \psi - \partial_z \phi e^{\phi} \bar{\psi} \partial_z \psi
- (\partial_z \phi)^2 e^{\phi} \bar{\psi} \psi\\
& \\
\Theta_4 = & - \partial_z \phi \partial_z^3 \phi e^{2 \phi} - \frac{17}{2}
(\partial_z \phi)^2 \partial_z^2 \phi e^{2 \phi} - 4 (\partial_z
\phi)^4 e^{2 \phi} + \frac{1}{2}\partial_z \phi \partial_z^3 \phi e^{\phi}
\bar{\psi} \psi +\\
& \\
& + \frac{11}{2} \partial_z \phi \partial_z^2 \phi e^{\phi} \bar{\psi}
\partial_z \psi - \frac{3}{2} (\partial_z \phi)^2 \partial_z^2 \phi e^{\phi}
\bar{\psi} \psi-\frac{21}{2} (\partial_z \phi)^2 e^{2 \phi} \psi 
\partial_z \psi +\\
& \\
& + \frac{3}{2} (\partial_z \phi)^3 e^{\phi} \bar{\psi} \partial_z
\psi + 3 (\partial_z \phi)^2 e^{\phi} \bar{\psi} \partial_z^2 \psi 
- \frac{11}{2} \partial_z \phi e^{2 \phi} \psi \partial_z^2 \psi - 
\frac{1}{2} e^{2 \phi} \psi \partial_z^3 \psi +\\
& \\
& + \frac{1}{2} \partial_z \phi e^{\phi} \bar{\psi} \partial_z^3 \psi 
- \frac{11}{2} \partial_z^2 \phi e^{2 \phi} \psi \partial_z \psi - 
 (\partial_z \phi)^4 e^{\phi} \bar{\psi} \psi
\end{array}
$$
These coincide with the results of ref.\cite{Chaichian}, which were
obtained by using B\"acklund transformations. Note that the system (10) is
invariant under $z \leftrightarrow \bar z$ and $\psi \to i
\bar{\psi}$, $\bar{\psi} \to i \psi$, so that there will be a
corresponding set of conserved quantities,
$$
\partial_z \bar T_{s+1} = \partial_{\bar z} \bar{\Theta}_{s-1}, \qquad
s=1,3,5,\cdots
$$
Even spin densities do not appear in the above sequence, because the
corresponding charges vanish. To see this, we note that eq.(8) remains
unchanged under the `gauge' transformation,
\begin{equation}
\left\{
\begin{array}{l}
J_1^{(n+1/2)} \to J_1^{(n+1/2)} + D_1 V^{(n+1/2)}\\
J_2^{(n+1/2)} \to J_2^{(n+1/2)} - D_2 V^{(n+1/2)}
\end{array}
\right.
\end{equation}
where $V^{(n+1/2)}$ is an arbitrary scalar superfield. It is then
straightforward to check that for the choices
$$
V^{(3/2)} = (\partial_- \varphi)^2 + 2i \theta_1 \partial_- \varphi
\partial_- \psi_2 + 4i \theta_2 \partial_- \varphi e^{2 \varphi}
\psi_2 - 4 \theta_1 \theta_2 e^{2 \varphi} (\psi_2 \partial_- \psi_2
+ i ( \partial_- \varphi)^2 )
$$
and 
$$
V^{(7/2)} =  2 (\partial_- \varphi)^4 + ( \partial_-^2 \varphi)^2 - 2i
\partial_- \psi_2 \partial_-^2 \psi_2 -16 i (\partial_- \varphi)^2
\psi_2 \partial_- \psi_2 +
$$
$$
 + i \theta_1 (16 (\partial_- \varphi)^2 \partial_-^2 \varphi \psi_2 +
2 \partial_-^3 \varphi \partial_- \psi_2 - 8( \partial_- \varphi)^3
\partial_- \psi_2) +
$$
$$
+ i \theta_2 e^{2 \varphi}( 8 \partial_- \varphi \partial_-^2 \varphi 
\psi_2 + 8 (\partial_- \varphi)^2 \partial_- \psi_2 + 4 \partial_-
\varphi \partial_-^2 \psi_2 - 16 ( \partial_- \varphi)^3 \psi_2) +
$$
$$
+ \theta_1 \theta_2 e^{2 \varphi} ( 16 i ( \partial_- \varphi)^4 - 36 i (
\partial_- \varphi)^2 \partial_-^2 \varphi - 8 \partial_-^2 \varphi
\psi_2 \partial_- \psi_2  + 64 ( \partial_- \varphi)^2 \psi_2
\partial_- \psi_2 + 4 \partial_- \psi_2 \partial_-^2 \partial_-^2 \psi_2),
$$ 
we get up to a transformation (11),
$$
\begin{array}{l}
J_A^{(5/2)} = \partial_- J_A^{(3/2)}\\
J_A^{(9/2)} = \partial_- J_A^{(7/2)}, \qquad (A=1,2)
\end{array}
$$
and the charges are thus trivial. This is consistent with the `spin
assignment' property discussed in ref. \cite{Bazhanov}. $T_2$, $\bar
T_2$ and $\Theta_0 = \bar{\Theta}_0$ are the components of the stress 
tensor. From the equations of motion it is easy to show that $\Theta_0$ is
just $\Theta_0 = \partial_z \partial_{\bar z} \phi$. We then have
$\partial_{\bar z} T = 0$, where,
\begin{equation}
T= T_2 - \partial_z^2 \phi = ( \partial_z \phi)^2 - \partial_z \psi
\psi - \partial_z^2 \phi.
\end{equation}
Similarly,
\begin{equation}
\bar T = ( \partial_{\bar z} \phi)^2 + \partial_{\bar z} \bar{\psi}
\bar{\psi} - \partial_{\bar z}^2 \phi.
\end{equation}
This is just the `conformally improved' stress tensor, \cite{Mansfield}. 
The total derivative terms restore the tracelessness of the stress
tensor, which is a necessary requirement for the theory to be
conformally invariant.

Besides being integrable, the theory on the full line is also
invariant under the supersymmetry transformations,
\begin{equation}
\left\{  
\begin{array}{l}
\delta_S \phi = \eta \psi + \bar{\eta} \bar{\psi}\\
\delta_S \psi = - ( \eta \partial_z \phi + \bar{\eta} e^{\phi} )\\
\delta_S \bar{\psi} = \bar{\eta} \partial_{\bar z} \phi + \eta
e^{\phi}
\end{array}
\right.
\end{equation}
where $\eta$ and $\bar{\eta}$ are infinitesimal constant fermionic 
parameters. To see this, let us first rewrite the action in Euclidean 
space after eliminating the non-dynamical auxiliary field F:
\begin{equation}
{\cal L}_0 = 2 ( \partial_z \phi \partial_{\bar z} \phi + \psi
\partial_{\bar z} \psi - \bar{\psi} \partial_z \bar{\psi} + e^{2 \phi}
- 2 e^{\phi} \bar{\psi} \psi).
\end{equation}
As expected the variation of ${\cal L}_0$ under these transformations 
then amounts to a total derivative:
\begin{equation}
\begin{array}{l l}
\delta_S {\cal L}_0 = & 2 \partial_z ( \delta_S \phi \partial_{\bar z}
\phi - \bar{\psi} \delta_S \bar{\psi} -2 \partial_{\bar z} \phi
\bar{\eta} \bar{\psi} - 2 \eta \bar{\psi} e^{\phi} ) +\\
& + 2 \partial_{\bar z} ( \delta_S \phi \partial_z \phi 
+ \psi \delta_S \psi -2 \partial_z \phi \eta \psi - 2 \bar{\eta} 
\psi e^{\phi}).
\end{array}
\end{equation}
The infinitesimal transformations (14) are generated by the currents
\begin{equation}
\left\{
\begin{array}{l}
J= \psi \partial_z \phi - \partial_z \psi\\
\bar J = \bar{\psi} \partial_{\bar z} \phi - \partial_
{\bar z} \bar{\psi} 
\end{array}
\right.
\end{equation}
These currents were also conformally improved by adding total
derivative terms by hand. The fact that a conformal improvement
is equally necessary for these currents is related to the fact that
they are supersymmetric partners of the stress tensor and are
therefore expected to have a similar behaviour, \cite{Mansfield}. We
note that this ensures that the theory be `chiral'. This also means
that we should be able to re-express the conservation laws derived
above in the form, $\partial_{\bar z} U_{s+1} = 0$. We have already
shown that this is true for $s=1$, where $U_2$ is the conformally
improved stress tensor (12). Similarly, we have 
$$
\Theta_2 = \partial_{\bar z} (2/3 ( \partial_z \phi)^3 + 
\partial_z \phi \psi \partial_z \psi),
$$
which means that $\partial_{\bar z} U_4 =0$, where
$$
U_4 = (\partial_z^2 \phi)^2 + ( \partial_z \phi)^4 + 3 ( \partial_z
\phi)^2 \psi \partial_z \psi + \partial_z \psi \partial_z^2 \psi - 2 (
\partial_z \phi)^2 \partial_z^2 \phi - \partial_z^2 \phi \psi
\partial_z \psi - \partial_z \phi \psi \partial_z^2 \psi.
$$
We can take the bosonic limit by setting the fermionic fields to
zero and we get $U_4^{(b)} = (T^{(b)})^2$, where $T^{(b)}$ is the
bosonic stress tensor $T^{(b)} = ( \partial_z \phi)^2 - \partial_z^2
\phi$.  This result motivates us to write $U_4 = T^2 + \delta U_4$, with:
$$
\delta U_4 = (\partial_z \phi)^2 \psi \partial_z \psi + \partial_z^2
\phi \psi \partial_z \psi + \partial_z \psi \partial_z^2 \psi -
\partial_z \phi \psi \partial_z^2 \psi.
$$
The two terms are separately conserved, i.e. $\partial_{\bar z} T^2 =
\partial_{\bar z} \delta U_4 = 0$. $\delta U_4$ should therefore be
expressible in terms of the supercurrent (17) and its
derivatives. Using dimensional condiderations, we obtained $\delta U_4
= J \partial_z J$, and so:
\begin{equation}
\left\{
\begin{array}{l}
U_4 = T^2 + J \partial_z J\\
\bar U_4 = \bar T^2 - \bar J \partial_{\bar z} \bar J
\end{array}
\right.
\end{equation}
For $s=5$, we get:
$$
\Theta_4 = \partial_{\bar z} \left [ \frac{3}{5} ( \partial_z \phi)^5
+ \partial_z \phi ( \partial_z^2 \phi)^2 - \frac{7}{2} ( \partial_z
\phi)^3 \partial_z^2 \phi - \frac{1}{2} \partial_z^2 \phi \partial_z^3
\phi - \frac{1}{2} \partial_z \phi \psi \partial_z^3 \psi + 
\right.
$$
$$
\left.
\qquad + \frac{1}{2} \partial_z^3 \phi \psi \partial_z \psi +
\frac{1}{2} \partial_z^2 \phi \psi \partial_z^2 \psi - \frac{13}{2}
\partial_z \phi \partial_z^2 \phi \psi \partial_z \psi + 2 (
\partial_z \phi)^3 \psi \partial_z \psi - 3 ( \partial_z \phi)^2 \psi
\partial_z^2 \psi \right].
$$
Using similar arguments, it is straightforward to show that:
\begin{equation}
\left\{
\begin{array}{l}
U_6 = T^3 + 1/2 ( \partial_z T)^2 + 2 TJ \partial_z J -
 1/2 J \partial_z^3 J\\
\bar U_6 = \bar T^3 + 1/2 ( \partial_{\bar z} \bar T)^2 - 2
 \bar T \bar J \partial_{\bar z} \bar J + 1/2 \bar J
 \partial_{\bar z}^3 \bar J
\end{array}
\right.
\end{equation}
We note that in the bosonic limit we recover exactly the IM of
 ref. \cite{Bazhanov}. 

\subsection{The $B^{(1)} (0,1)$ theory}

The $B^{(1)} (0,1)$ theory is defined by the superspace equation,
\begin{equation}
D_1D_2 \Phi = i e^{2 \Phi} - \frac{1}{2} \theta_1 \theta_2 e^{-4
\Phi}.
\end{equation}
The second term on the right-hand side spoils invariance under
supersymmetry. This is a common feature of Toda theories based on
contragradient Lie superalgebras, \cite{Evans} \cite{Olshanetsky}.

Alternatively, eq.(20) can be seen as the compatibilty condition for
a linear system similar to (1), where this time the graded matrices
take the form:
$$
A_1 (\lambda) = 
\left(
\begin{array}{c c c}
-2 D_1 \Phi & -i \lambda \sqrt 2 & 0\\
0 & 0 & i \lambda \sqrt 2\\
- \lambda \theta_2 & 0 & 2 D_1 \Phi
\end{array}
\right)
\qquad 
A_2 (\lambda) = \frac{1}{\lambda} 
\left(
\begin{array}{c c c}
0 & 0 & \theta_1 e^{-4 \Phi}\\
\sqrt 2 e^{2 \Phi} & 0 & 0\\
0 & \sqrt 2 e^{2 \Phi} & 0
\end{array}
\right).
$$
Expressing eq.(20) in components, we get in Euclidean space:
\begin{equation}
\left\{
\begin{array}{l}
F=-i e^{\phi}; \partial_z \bar{\psi} = - e^{\phi} \psi ; \partial_{\bar
z} \psi = - e^{\phi} \bar{\psi}\\
\partial_z \partial_{\bar z} \phi = e^{2 \phi} - e^{\phi} \bar{\psi}
\psi - \frac{1}{4} e^{-2 \phi}
\end{array}
\right.
\end{equation}
The bosonic limit of this theory is the $a_1^{(1)}$ bosonic Toda
theory. It was conjectured \cite{Olshanetsky} that the gaps in the
sequence of conservation laws be periodic with period equal to
2. Specificaly, there will be an infinite set of conserved densities,
$\partial_{\bar z} T_{s+1} = \partial_z \Theta_{s-1}$, with
$s=1,3,5,\cdots$ 

Considering the most general Ansatz, we obtained the following
elements:
\begin{equation}
\begin{array}{l l}
T_2 = & (\partial_z \phi)^2 - \partial_z \psi \psi\\
T_4 = & (\partial_z^2 \phi)^2 + (\partial_z \phi)^4 + 3 ( \partial_z
\phi)^2 \psi \partial_z \psi + \partial_z \psi \partial_z^2 \psi + 3
\partial_z \phi \partial_z^2 \psi \psi\\
& \\
\Theta_0 = &e^{2 \phi} -e^{\phi} \bar{\psi} \psi + \frac{1}{4} e^{-2
\phi} \\ 
 \Theta_2 = & 2 (\partial_z \phi)^2 e^{2 \phi} + 4 (\partial_z \phi)^2
e^{\phi} \psi \bar{\psi} + \frac{1}{2} ( \partial_z \phi)^2 e^{-2
\phi} + 2 e^{2 \phi} \partial_z \psi \psi + \frac{3}{2} e^{-2 \phi}
\psi \partial_z \psi + \\
 & +2 \partial_z \phi e^{\phi} \bar{\psi} \partial_z \psi
\end{array}
\end{equation}

\section{The theories on the half-line}

Let us assume a boundary located at $x=0$. We will follow closely
ref. \cite{Inami}.

The action on the half-line $x \in (- \infty , 0]$ is the sum of two
contributions
$$
S=S_0 + S_{{\cal B}} \equiv \int_{- \infty}^{+ \infty} dx \int_{-
\infty}^{+ \infty} dy \left\{ \theta (-x) {\cal L}_0 + \delta (x)
{\cal B} (\phi, \psi, \bar{\psi}) \right\},
$$
where ${\cal L}_0$ is the bulk lagrangian density for either theory 
and the boundary potential ${\cal B}$ is assumed to be independent 
of the field derivatives. $\theta$ is the Heaviside step function.

Minimizing the action leads to the bulk field equations. Furthermore, we get
the boundary conditions at $x=0$:
\begin{equation}
\partial_x \phi + \frac{\partial {\cal B}}{\partial \phi} = 0, \qquad
\psi - \frac{\partial {\cal B}}{\partial \psi} = 0, \qquad \bar{\psi}
+ \frac{\partial {\cal B}}{\partial \bar{\psi}}=0.
\end{equation}

\subsection{The super-Liouville theory}

We first investigate under what circumstances supersymmetry will be
preserved in the presence of a boundary. It turns out, as we shall
see, that it is still possible to keep half of the supersymmetries,
provided one chooses suitable boundary conditions. These conditions
equally preserve the conformal invariance of the theory and therefore
its integrability.

Let us start by writing explicitly the variation of the action eq.(16)
under the transformations (14) in the presence of the boundary.
$$
\delta_S S_0 = \int_{-\infty}^{+ \infty} dy \left\{ \frac{1}{2} ( \eta
\psi + \bar{\eta} \bar{\psi}) \phi_x + ( \bar{\psi} \eta + \psi
\bar{\eta}) e^{\phi} + \frac{i}{2} ( \eta \psi - \bar{\eta} \bar{\psi})
\phi_y \right\} |_{x=0}.
$$
This expression can be compensated for by adding a boundary term. On
dimensional grounds, we consider a boundary potential of the form:
$$
{\cal B_S} = c_S e^{\phi} + M_S \bar{\psi} \psi.
$$
The boundary equations of motion arising from this term are:
$$
\phi_x = - c_S e^{\phi}, \qquad \psi + M_S \bar{\psi} =0, \qquad
(M_S^2 =1).
$$
Under a supersymmetry transformation, we have:
$$
\begin{array}{c}
\delta_S \int_{- \infty}^{+ \infty} dy {\cal B_S} = \\
= \int_{ - \infty}^{
+ \infty} dy \left\{ \frac{1}{2} M_S ( \bar{\eta} \psi + \eta
\bar{\psi} ) \phi_x + ( c_S + M_S ) ( \eta \psi + \bar{\eta}
\bar{\psi} ) e^{\phi} + \frac{i}{2} M_S ( \bar{\eta} \psi + \bar{\psi}
\eta ) \phi_y \right\}.
\end{array}
$$
It is only possible to keep half of the supersymmetries. We therefore
choose $ \bar{\eta} = \pm \eta$. The sum of the two contributions is
thus:
$$
\begin{array}{c}
\delta_S S_0 + \delta_S S_{{\cal B_S}} = \\
= \int_{- \infty}^{+ \infty} dy \left\{ \frac{1}{2} ( 1 \pm M_S) \eta
(\psi \pm \bar{\psi}) \phi_x + (c_S + M_S \mp 1) \eta ( \psi \pm
\bar{\psi}) e^{\phi} + \frac{i}{2} ( 1 \pm M_S) \eta ( \psi \mp
\bar{\psi}) \phi_y \right\}.
\end{array}
$$
The integrand in the above expression will be a total y-derivative, if
we choose one of the following possibilities:
$$
\begin{array}{l l l}
1) & \phi_x = \mp 2 e^{\phi}, & \psi \mp \bar{\psi}=0\\
2) & \phi_x = \mp 2 e^{\phi}, & \psi_y=\bar{\psi}_y=0\\
3) & \phi_x =- c_S e^{\phi} , & \psi \pm \bar{\psi} =0, \qquad 
\phi_y =0
\end{array}
$$
It is easy to show that under an infinitesimal supersymmetry
transformation,
$$
\begin{array}{l l}
\delta_S(\phi_x + c_S e^{\phi}) = -i \eta \partial_y ( \psi \mp
\bar{\psi}) + (c_S \mp 2)\eta (\psi \pm \bar{\psi}) e^{\phi},
& \delta_S(\phi_y) = \eta \partial_y ( \psi \pm \bar{\psi}),\\
\delta_S (\psi_y) = -\frac{\eta}{2} \partial_y ( \phi_x \pm 2
e^{\phi} - i \phi_y) , & \delta_S( \bar{\psi}_y) = \pm \frac{\eta}{2} 
\partial_y (\phi_x \pm 2 e^{\phi} + i \phi_y),\\
\delta_S (\psi  \pm \bar{\psi}) = i \eta \phi_y, & \delta_S ( \psi \mp
\bar{\psi}) = - \eta ( \phi_x \pm 2 e^{\phi}).
\end{array}
$$
If in 2) we take $\phi_y=0$, the first two cases will be supersymmetry
preserving, whilst the latter will not. 

Let us consider additional terms in the boundary potential, 
$\epsilon_S \psi + \bar{\epsilon}_S \bar{\psi}$.
Under a supersymmetry transformation such that $\bar{\eta} = \pm
\eta$:
$$
\delta_S( \epsilon_S \psi + \bar{\epsilon}_S \bar{\psi} ) = -
\frac{1}{2} ( \epsilon_S \mp \bar{\epsilon}_S) \eta \phi_x \mp (
\epsilon_S \mp \bar{\epsilon}_S) \eta e^{\phi} + \frac{i}{2} (
\epsilon_S \pm \bar{\epsilon}_S) \eta \phi_y.
$$
This will be a total y-derivative if
$$
\begin{array}{l}
1) \qquad \epsilon_S \mp \bar{\epsilon}_S = 0\\
2) \qquad \epsilon_S \pm \bar{\epsilon}_S = 0, \qquad \phi_x = \mp 2
e^{\phi}
\end{array}
$$
In summary, the boundary potential,
\begin{equation}
{\cal B_S} = \pm 2 e^{\phi} + M_S \bar{\psi} \psi + \epsilon_S \psi +
\bar{\epsilon}_S \bar{\psi},
\end{equation}
restores supersymmetry, provided:
$$
\begin{array}{l l l l}
1) & M_S = \mp 1, & \bar{\epsilon}_S = \pm \epsilon_S, & \psi \mp
\bar{\psi} = - \epsilon_S\\
2) & M_S^2 \ne 1, & \bar{\epsilon}_S = \mp \epsilon_S, & \phi_y = 0\\
& & \psi = - \frac{\epsilon_s}{1 \mp M_S}, & \bar{\psi} = \frac{\pm
\epsilon}{1 \mp M_S}  
\end{array}
$$
Let us now discuss the integrability of the theory. According to
Cardy, \cite{Cardy}, invariance of the boundary conditions under a
symmetry generated by some set of conserved currents $(W^{(r)}, \bar
W^{(r)})$ requires $W^{(r)} = \bar W^{(r)}$ on the boundary.
If we take $W^{(r)}$ to be the stress tensor, we get from eqs.(12), (13):
$$
-i \phi_x \phi_y - \frac{1}{2} \psi_x \psi + \frac{i}{2} \psi_y \psi +
i \phi_{xy} = \frac{1}{2} \bar{\psi}_x \bar{\psi} + \frac{i}{2}
\bar{\psi}_y \bar{\psi}.
$$
The bosonic part, $-i\phi_x \phi_y + i\phi_{xy}$, vanishes for $\phi_x = c
e^{\phi}$, in which case :
\begin{equation}
\psi_y \psi = \bar{\psi}_y \bar{\psi},
\end{equation}
where we used the equations of motion (10) to eliminate the
x-derivatives. There are two solutions to eq.(25):
\begin{equation}
\begin{array}{l l}
1) & \bar{\psi} = \pm \psi\\
2) & \psi_y = \bar{\psi}_y = 0
\end{array}
\end{equation}
We notice that these solutions are reminescent of the conditions
obtained for the conservation of supersymmetry. However, $c$ and
$\phi_y$ remain arbitrary. But we can still impose similar constraints
on the supercurrents $(J, \bar J)$ and this should fix $c$ and
$\phi_y$. The boundary condition is $\bar{\eta} \bar{\psi} = \eta
\psi$. Remember that we want to keep half of the supersymmetries, by 
setting $ \bar{\eta} = \pm \eta$. Accordingly, we impose the boundary 
condition $\bar J = \mp J$ and get:
\begin{equation}
(\phi_x + i \phi_y \pm 2 e^{\phi} ) \bar{\psi} - 2i \bar{\psi}_y = \mp
( \phi_x - i \phi_y \pm 2 e^{\phi} ) \psi \mp 2i \psi_y,
\end{equation}
where we used the explicit expressions (17) for $J$ and $\bar J$. Let
us use as Ansatz the conditions (26) obtained above for the conservation of
conformal invariance. We then get:
\begin{equation}
\begin{array}{l l l}
1) & \bar{\psi} = \pm \psi, & \phi_x = \mp 2 e^{\phi}\\
2) & \psi_y = \bar{\psi}_y = 0, & \phi_x = \mp 2 e^{\phi}, \qquad
\phi_y =0
\end{array}
\end{equation}
In 1) the sign of $\bar{\psi}$ was chosen so as to cancel the terms
proportional to $\psi_y$ and $\bar{\psi}_y$ in eq.(27). It is easy to 
check that these conditions also preserve the following combinations of the IM:
$$
I_s = \int_{- \infty}^0 dx ( U_{s+1} + \bar U_{s+1}), \qquad (s=1,3,5,\cdots).
$$ 
We just have to show that $U_{s+1} = \bar U_{s+1}$ at $x=0$ as a
consequence of the stress tensor and the supercurrent satisfying $T=
\bar T$ and $\bar J = \mp J$.

All polynomials $T^n$ ($n>1$) automatically satisfy $T^n = \bar
T^n$. The first non-trivial term is $(\partial_z T)^2$. From the 
conservation of the stress tensor, we have $T_x = - i T_y$, $\bar T_x
= i \bar T_y$. This implies that at $x=0$,
$$
(\partial_z T)^2 = - T_y^2 = - \bar T_y^2 = (\partial_{\bar z} \bar
T)^2.
$$
Similarly, from $J_x = -i J_y$, $\bar J_x = i \bar J_y$, we have:
$$
J \partial_z J = - i J J_y = - i ( \mp \bar J)(\mp \bar J_y) = - i
\bar J \bar J_y = - \bar J \partial_{\bar z} \bar J.
$$
Altogether, this means that $U_4 = \bar U_4$. Next we consider the
term $J \partial_z^3 J$. We use the following identities,
$$
\left\{
\begin{array}{l r r r}
J_{xyy} = &i J_{xxy} = & - J_{xxx} = & -i J_{yyy}\\
\bar J_{xyy} = & - i \bar J_{xxy} = & - \bar J_{xxx} = & i \bar
J_{yyy}
\end{array}
\right.
$$
to show that 
$$
J \partial_z^3 J = \frac{1}{8} J ( J_{xxx} - 3 i J_{xxy} - 3 J_{xyy} + i
J_{yyy}) = i J J_{yyy} = i \bar J \bar J_{yyy} = - \bar J
\partial_{\bar z}^3 \bar J.
$$
Again, we have $U_6 = \bar U_6$. As advertised, the conditions (28) 
coincide precisely with the conditions for conservation of
supersymmetry, provided we take $\epsilon_S=0$. In summary, we found 
that the conditions preserving half of the supersymmetries in the 
surface configuration, also ensure the conservation of the
superconformal invariance and the IM. This is
not surprising, since the IM, being composite fields of the stress
tensor, the supercurrents and their derivatives, are deeply connected 
with the superconformal symmetry of the theory.

\subsection{The $B^{(1)} (0,1) $ theory}

Suppose that the boundary potential ${\cal B}$ can be chosen in such a
way that at $x=0$
\begin{equation}
T_{s+1} - \bar T_{s+1} - ( \Theta_{s-1} - \bar{\Theta}_{s-1}) =
\frac{d}{dy} \Sigma_s (y),
\end{equation}
where $\Sigma_s (y)$ is some functional of the boundary fields. Then
the `spin' s charge given by
$$ 
Q_s = \int_{- \infty}^0 dx ( T_{s+1} + \bar T_{s+1} + \Theta_{s-1} +
\bar{\Theta}_{s-1} ) - i \Sigma_s (y)
$$
is a non-trivial IM, \cite{Ghoshal}. Let us now look for potentials
that produce expressions like eq.(29). From the equations of motion, 
we have:
$$
\left\{
\begin{array}{l}
\psi_{xy} = -2 e^{\phi} \bar{\psi}_y - 2 \phi_y e^{\phi} \bar{\psi} -
i \psi_{yy}\\
\psi_{xx} = 4 e^{2 \phi} \psi -2 \phi_x e^{\phi} \bar{\psi} + 2i
\phi_y e^{\phi} \bar{\psi} - \psi_{yy}\\
\bar{\psi}_{xy} = - 2 e^{\phi} \psi_y - 2 \phi_y e^{\phi} \psi + i
\bar{\psi}_{yy}\\
\bar{\psi}_{xx} = 4 e^{2 \phi} \bar{\psi} - 2 \phi_x e^{\phi} \psi -
2i \phi_y e^{\phi} \psi - \bar{\psi}_{yy}\\
\end{array}
\right.
$$
$$
\phi_{xx} = 4 e^{2 \phi} - \phi_{yy} - 4 e^{\phi} \bar{\psi} \psi -
e^{-2 \phi}.
$$
Moreover, from the eq.(23) , we have at $x=0$:
$$
\frac{\partial^2 {\cal B}}{\partial \phi \partial \psi} =
\frac{\partial^2 {\cal B}}{\partial \phi \partial \bar{\psi}} = \frac{
\partial^2 {\cal B}}{\partial \psi \partial \bar{\psi}} = 0.
$$
Consequently,
$$
\phi_{xy} = - \frac{\partial^2 {\cal B}}{\partial \phi^2} \phi_y.
$$
Using these expressions, we get:
$$
T_4 - \bar T_4 + \bar{\Theta}_2 - \Theta_2 = {\cal W}_b + {\cal W}_R,
$$
where ${\cal W}_b$ is a purely bosonic contribution,
$$
{\cal W}_b = - \frac{i}{4} \frac{\partial^2 {\cal B}}{\partial \phi^2}
\phi_{yy} \phi_y - \frac{i}{8} \frac{\partial {\cal B}}{\partial \phi}
\phi_y^3 + \frac{i}{2} \left\{ \frac{1}{4} \left( \frac{\partial {\cal
B}}{\partial \phi} \right)^3 + \frac{\partial^ 2 {\cal B}}{\partial
\phi^2} \left( e^{2 \phi} - \frac{1}{4} e^{-2 \phi} \right)
-\frac{\partial {\cal B}}{\partial \phi} \left( e^{2 \phi} +
\frac{1}{4} e^{-2 \phi} \right) \right\}\phi_y.
$$
We look for solutions of the form ${\cal B} (\phi, \psi , \bar{\psi})
= {\cal B}_b (\phi) + {\cal B}_f ( \psi, \bar{\psi})$. It is
straightforward to show that for ${\cal B}_b (\phi) = a e^{\phi} + b
e^{- \phi}$, where $a$ and $b$ are arbitrary constants, ${\cal W}_b$ will
automatically be a total y-derivative. The remaining contribution
${\cal W}_R$ is given by:
$$
\begin{array}{c}
{\cal W}_R = - \frac{i}{2} \bar{\psi} \psi \phi_y
\left(\frac{\partial^2 {\cal B}_b}{\partial \phi^2} + \frac{\partial
{\cal B}_b}{\partial \psi} \right) e^{\phi} - \frac{3i}{8}
\left[\left( \frac{\partial {\cal B}_b}{\partial \phi} \right)^2 -
\phi_y^2 - 4e^{2 \phi} - e^{- 2 \phi} \right] ( \bar{\psi}_y
\bar{\psi} - \psi_y \psi)+\\
+ \frac{3}{4} \frac{\partial {\cal B}_b}{\partial \phi} \phi_y (
\bar{\psi} \psi_y + \psi \bar{\psi}_y) + \frac{i}{2} \frac{\partial 
{\cal B}_b}{\partial \phi} e^{\phi} (\psi_y \bar{\psi} - 
\bar{\psi}_y \psi) + 2 \phi_y e^{\phi} ( \psi_y \bar{\psi} +
\bar{\psi}_y \psi)+\\
+ i(\bar{\psi}_{yy} \bar{\psi}_y - \psi_{yy} \psi_y) + e^{\phi} (
\bar{\psi} \psi_{yy} - \bar{\psi}_{yy} \psi) + \frac{3}{4}
\frac{\partial {\cal B}_b}{\partial \phi} ( \psi_{yy} \psi +
\bar{\psi}_{yy} \bar{\psi}) + \frac{3i}{4} \phi_y ( \psi_{yy} \psi -
\bar{\psi}_{yy} \bar{\psi}).
\end{array}
$$
Because $\psi$, $\bar{\psi}$ are Grassmann variables, ${\cal B}_f$
takes the form ${\cal B}_f ( \psi, \bar{\psi}) = M \bar{\psi} \psi +
\epsilon \psi + \bar{\epsilon} \bar{\psi}$, where $M$, $\epsilon$,
$\bar{\epsilon}$ are constant parameters, $M$ being bosonic and the
remaining fermionic. From eq.(23), we have the following possibilities
at $x=0$:
$$
\begin{array}{l}
1) \qquad \psi = - \frac{ \epsilon + M \bar{\epsilon}}{1- M^2}, \qquad
\bar{\psi} = \frac{\bar{\epsilon} + M \epsilon}{1-M^2}, \qquad M \ne
\pm 1\\
2) \qquad \bar{\psi} = \mp ( \psi + \epsilon), \qquad \bar{\epsilon} =
\mp \epsilon, \qquad M = \pm 1
\end{array}
$$
In the first case, ${\cal W}_R$ is automatically a total y-derivative,
irrespective of the values of $a,b,\epsilon, \bar{\epsilon}$ and
$M(\ne \pm 1)$. In the latter case, we get $\epsilon = \bar{\epsilon}
=0$ and $a=\mp2$, corresponding to $M=\pm 1$. In summary, there will
be a spin $s=3$ conserved charge in the following cases:
$$
\begin{array}{l l}
1) & {\cal B} ( \phi, \psi, \bar{\psi}) = a e^{\phi} + b
e^{-\phi} + M \bar{\psi} \psi + \epsilon \psi + \bar{\epsilon}
\bar{\psi} \\
& \phi_x = -a e^{\phi} + b e^{- \phi} , \qquad \psi = 
- \frac{ \epsilon + M \bar{\epsilon}}{1- M^2}, \qquad
\bar{\psi} = \frac{\bar{\epsilon} + M \epsilon}{1-M^2}, \\
& \qquad a,b,\epsilon, \bar{\epsilon} $ and $ M (\ne
\pm 1) $ are arbitrary.$\\
2) & {\cal B} ( \phi, \psi, \bar{\psi}) =  \mp 2 e^{\phi} + b
e^{-\phi} \pm \bar{\psi} \psi\\
& \phi_x = \pm 2 e^{\phi} + b e^{- \phi}, \qquad \psi \pm \bar{\psi} =
0\\
& \qquad b$ is arbitrary.$
\end{array}
$$

\section{Conclusions}

Let us restate our results. We derived a recursive formula for an
infinity of integrals of motion (IM) for the super-Liouville-equation
(SLE) which consist of a supersymmetric extension of the classical 
expressions in ref.\cite{Bazhanov}. These IM and its eigenstates, 
together with the factorisable S-matrix constitute an alternative 
description of the conformal theory. 

The boundary equations of motion, preserving half of the supersymmetries,
automatically conserve the superconformal invariance and half of the
IM on the half-line. Furthermore, these boundary conditions are
unambiguously determined, i.e. there are no unfixed parameters. A
similar situation occurs in the super-sine-Gordon theory \cite{Inami}
and appears to be a consequence of supersymmetry. Indeed, our analysis
of the non-supersymmetric $B^{(1)} (0,1)$ theory reveals that, in
contrast with the two models above, the boundary potential depends on
free parameters.

\begin{center}
{\large {\bf Acknowledgements}}
\end{center}

The author wishes to thank A.Taormina, P.Bowcock and E.Corrigan for
useful discussions and for reading the manuscript.

This work is supported by J.N.I.C.T.'s PRAXIS XXI PhD fellowship 
BD/3922/94.

\end{document}